\documentclass[12pt]{article}

\usepackage{amsmath}
\usepackage{graphicx}

\newcommand{\ket}[1]{| #1 \rangle}
\newcommand{\bra}[1]{\langle #1 |}
\newcommand{\hcs}[1]{#1^\dagger #1}
\newcommand{\expv}[1]{\langle #1 \rangle}

\begin{document}

\begin{center}
{\Large\bf Information, fidelity, and reversibility
in single-qubit measurements}
\vskip .6 cm
Hiroaki Terashima
\vskip .4 cm
{\it Department of Physics, Faculty of Education, Gunma University, \\
Maebashi, Gunma 371-8510, Japan}
\vskip .6 cm
\end{center}

\begin{abstract}
We explicitly calculate information, fidelity,
and reversibility of an arbitrary single-qubit measurement
on a completely unknown state.
These quantities are expressed as functions of a single parameter,
which is the ratio of the two singular values of the measurement operator
corresponding to the obtained outcome.
Thus, our results give information tradeoff relations
to the fidelity and to the reversibility
at the level of a single outcome rather than
that of an overall outcome average.
\end{abstract}

\begin{flushleft}
{\footnotesize
{\bf PACS}: 03.65.Ta, 03.67.-a\\
{\bf Keywords}: quantum measurement, quantum information
}
\end{flushleft}

\section{Introduction}
Quantum measurement provides information on a physical system,
while it inevitably changes the state of the system
depending on the obtained outcome.
This property is of great interest
in the foundations of quantum mechanics and
is of practical importance in quantum information processing
and communication~\cite{NieChu00}
such as quantum cryptography~\cite{BenBra84,Ekert91,Bennet92,BeBrMe92}.
Therefore, 
numerous studies~\cite{FucPer96,Banasz01,FucJac01,BanDev01,DArian03,Ozawa04,%
Maccon06,Sacchi06,BusSac06,BuHaHo07}
have discussed tradeoff relations between
the information gain and the state change in quantum measurement
by quantifying them in various ways.
For example, Banaszek~\cite{Banasz01} has shown
an inequality between two fidelities quantifying
the information gain and the state change.

Interestingly, in connection with such a state change,
quantum measurement was widely believed
to have intrinsic irreversibility~\cite{LanLif77}
because of non-unitary state reduction.
However, it has been shown that
quantum measurement is not necessarily
irreversible~\cite{UedKit92,UeImNa96}
if all the information on the system is preserved
during the measurement process.
In particular, a quantum measurement is said to
be physically reversible~\cite{UeImNa96,Ueda97}
if the pre-measurement state can be recovered from
the post-measurement state
with a non-zero probability of success
by means of a second measurement,
known as reversing measurement.
Several physically reversible measurements have been proposed with
various systems~\cite{Imamog93,Royer94,TerUed05,KorJor06,%
TerUed07,SuAlZu09,XuZho10}
and have been experimentally demonstrated
using various qubits~\cite{KNABHL08,KCRK09}.
Thus, it would be interesting
to involve physical reversibility
while discussing information tradeoff relations.
In fact, a recent discussion
on photodetection processes~\cite{Terash10} has suggested
the existence of a tradeoff relation between the information gain
and the physical reversibility.
Such a tradeoff relation is also expected in view of
a different type of reversible measurement,
known as unitarily reversible measurement~\cite{MabZol96,NieCav97},
in which the pre-measurement state can be recovered with unit probability
by means of a unitary operation,
whereas the measurement provides no information
about the measured system.

Moreover, physically reversible measurements
naturally prompt investigation of the information tradeoff relation
at the level of a single outcome~\cite{DArian03}
rather than that of an overall outcome average
because the state recovery by
reversing measurement relies on the postselection of outcomes.
That is, the reversing measurement
can recover the state of the system changed by
a physically reversible measurement
only when it yields a preferred outcome.
Unfortunately,
this state recovery is always accompanied by
the erasure of information 
obtained by the physically reversible measurement
(Erratum of \cite{Royer94}),
implying a tradeoff relation between the information gain and
the state change at the single outcome level.
However, an approximate recovery by
the Hermitian conjugate measurement~\cite{TerUed07b}
does not necessarily decrease the information gain.

In this paper,
we derive general formulae for the
information gain, the state change, and the physical reversibility
in quantum measurements,
in which the system to be measured is a two-level system or qubit
in a completely unknown state.
We evaluate the amount of information gain
by using a decrease in Shannon entropy~\cite{DArian03,TerUed07b},
the degree of state change by using fidelity~\cite{Uhlman76},
and the degree of physical reversibility by
using the maximal successful probability
of reversing measurement~\cite{KoaUed99}.
Because the formulae are written as functions of a single parameter,
they lead to information tradeoff relations
to the state change and the physical reversibility
at a single outcome level.
We also consider two efficiencies of the measurement
with respect to the state change and the physical reversibility,
and we show their different behaviors as functions of the single parameter.

This paper is organized as follows:
Section~\ref{sec:formulation} explains the procedure to quantify
the information gain, the state change, and the physical reversibility,
and it shows their explicitly calculated formulae in the case of
an arbitrary single-qubit measurement.
Section~\ref{sec:tradeoff} discusses
information tradeoff relations
to the state change and the physical reversibility,
and it defines two efficiencies of the measurement
with respect to
the state change and the physical reversibility.
Section~\ref{sec:conclude} summarizes our results.

\section{\label{sec:formulation}Formulation}


To evaluate the amount of information provided by a single-qubit measurement,
we assume that the pre-measurement state of the qubit
is known to be one of the predefined pure states $\{\ket{\psi(a)}\}$
with equal probability, $p(a)=1/N$, where $a=1,\ldots,N$,
although the index $a$ of the pre-measurement state is unknown to us.
Since the pre-measurement state is usually
an arbitrary unknown state in quantum measurement,
the set $\{\ket{\psi(a)}\}$ actually consists of
all possible pure states of the qubit with $N\to\infty$.
The lack of information on the state of the qubit
can initially be evaluated by the Shannon entropy as
\begin{equation}
  H_0=-\sum_a p(a)\log_2 p(a)=\log_2 N.
\label{eq:h0}
\end{equation}

Next, we measure the qubit
to obtain information on its state.
In a more general formulation of
quantum measurement~\cite{DavLew70,NieChu00},
a quantum measurement is described by a set of
measurement operators $\{\hat{M}_m\}$ that satisfies
\begin{equation}
\sum_m\hcs{\hat{M}_m}=\hat{I},
\end{equation}
where $\hat{I}$ is the identity operator.
That is, if the system to be measured is in a state $\ket{\psi}$,
the measurement yields an outcome $m$ with probability
\begin{equation}
 p_m=\bra{\psi}\hcs{\hat{M}_m}\ket{\psi},
\label{eq:probability}
\end{equation}
causing a state reduction of the measured system to
\begin{equation}
\ket{\psi_m}=\frac{1}{\sqrt{p_m}}\hat{M}_m\ket{\psi}.
\label{eq:reduction}
\end{equation}
Here, we have assumed that the quantum measurement
is efficient~\cite{FucJac01} or ideal~\cite{NieCav97}
to ignore classical noise that yields a mixed post-measurement state,
because we are interested in the quantum nature of measurement.
From now on, we focus on a single measurement process
with outcome $m$
described by a measurement operator $\hat{M}_m$.
The measurement operator $\hat{M}_m$
can always be written by singular-value decomposition as
\begin{equation}
 \hat{M}_m=\kappa_m\hat{U}_m\hat{D}_m\hat{V}_m,
\end{equation}
where $\kappa_m$ is a real number,
$\hat{U}_m$ and $\hat{V}_m$ are unitary operators,
and $\hat{D}_m$ is a non-negative operator
with diagonal matrix representation
in an orthonormal basis $\{\ket{0},\ket{1}\}$,
\begin{equation}
\hat{D}_m=\ket{0}\bra{0}+\lambda_m\ket{1}\bra{1}
   =\left(\begin{array}{cc}
    1 & 0 \\
    0 & \lambda_m
    \end{array}\right)
\end{equation}
with $0\le\lambda_m\le1$, for the single-qubit measurement.
Note that the diagonal element $\lambda_m$ is
the ratio of the two singular values of $\hat{M}_m$.
Without loss of generality,
we can omit the unitary operator $\hat{V}_m$ as
\begin{equation}
 \hat{M}_m=\kappa_m\hat{U}_m\hat{D}_m,
\label{eq:operator}
\end{equation}
by relabeling the index $a$ as
$\ket{\psi'(a)}=\hat{V}_m\ket{\psi(a)}$.

If the pre-measurement state is $\ket{\psi(a)}$,
measurement (\ref{eq:operator})
yields the outcome $m$ with probability
\begin{equation}
  p(m|a) = \kappa_m^2\bra{\psi(a)}\hat{D}_m^2 \ket{\psi(a)}
        \equiv \kappa_m^2q_m(a)
\label{eq:prob}
\end{equation}
as given in Eq.~(\ref{eq:probability}).
Since the probability for $\ket{\psi(a)}$ is $p(a)=1/N$,
the total probability for the outcome $m$ is given by
\begin{equation}
  p(m) =\sum_a  p(m|a)\,p(a)=\frac{1}{N}\sum_a \kappa_m^2q_m(a)=
     \kappa_m^2\overline{q_m},
\label{eq:totalprob}
\end{equation}
where the overline denotes the average over $a$,
\begin{equation}
   \overline{f} \equiv \frac{1}{N}\sum_a f(a).
\end{equation}
On the contrary,
given the outcome $m$, we can find the probability for
the pre-measurement state $\ket{\psi(a)}$ as
\begin{equation}
  p(a|m) =\frac{p(m|a)\,p(a)}{p(m)}=\frac{q_m(a)}{N\,\overline{q_m}}
\label{eq:conditional}
\end{equation}
from Bayes' rule,
which means that the lack of information on the pre-measurement state
becomes the Shannon entropy
\begin{equation}
  H(m) =-\sum_a p(a|m)\log_2 p(a|m)
\end{equation}
after the measurement.
Therefore, the information gain by the measurement with
the \emph{single} outcome $m$ can be defined by the decrease
in Shannon entropy as~\cite{DArian03,TerUed07b}
\begin{equation}
  I(m) \equiv H_0-H(m)
    =\frac{\overline{q_m\log_2 q_m} -\overline{q_m}\log_2 \overline{q_m}}
     {\overline{q_m}}.
\label{eq:info}
\end{equation}
Note that this information gain is positive and is free from
the divergent term $\log_2 N\to\infty$ in Eq.~(\ref{eq:h0}).
These results essentially arise from the assumption that
the probability distribution $p(a)$ is uniform.
If averaged over all the outcomes, the information gain
reduces to the mutual information~\cite{NieChu00}
of the random variables $\{a\}$ and $\{m\}$, namely,
\begin{equation}
   I\equiv\sum_m p(m)\,I(m)=
   \sum_{m,a} p(a|m)\,p(m)\, \log_2\frac{p(a|m)}{p(a)}.
\label{eq:av_info}
\end{equation}

\begin{figure}
\begin{center}
\includegraphics[scale=0.6]{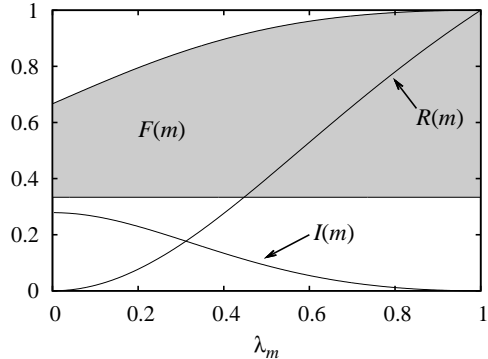}
\end{center}
\caption{\label{fig1}
Information gain $I(m)$, fidelity $F(m)$, and reversibility $R(m)$
when the measurement yields a single outcome $m$,
as functions of $\lambda_m$.
The parameter $\lambda_m=0$ corresponds to a projective measurement,
and $\lambda_m=1$ corresponds to the identity operation
except for a unitary operation.}
\end{figure}

To explicitly calculate the information gain (\ref{eq:info}),
we parameterize the state of the qubit by
two continuous angles $(\theta,\phi)$ as
\begin{equation}
  \ket{\psi(a)}=\cos\frac{\theta}{2}\,\ket{0}+
    e^{i\phi}\sin\frac{\theta}{2}\,\ket{1},
\end{equation}
where $0\le\theta\le\pi$ and $0\le\phi<2\pi$.
Thus, the summation over $a$ is replaced
with an integral over $(\theta,\phi)$ as
\begin{equation}
   \frac{1}{N}\sum_a   \quad\longrightarrow\quad
     \frac{1}{4\pi}\int^{2\pi}_0 d\phi\,
     \int^\pi_0 d\theta \sin \theta.
\end{equation}
Since
\begin{equation}
  q_m(a)=\cos^2\frac{\theta}{2}+\lambda_m^2\sin^2\frac{\theta}{2}
\end{equation}
from Eq.~(\ref{eq:prob}),
the information gain (\ref{eq:info}) is calculated to be
\begin{equation}
I(m) =1-\frac{1}{2\ln2}-\frac{\lambda_m^4}{1-\lambda_m^4}\log_2\lambda_m^2
           -\log_2\left(1+\lambda_m^2\right),
\label{eq:information}
\end{equation}
which depends only on $\lambda_m$.
Figure~\ref{fig1} shows the information gain $I(m)$
as a function of $\lambda_m$.
The information gain $I(m)$ has a maximal value
$1-1/(2\ln2)$ at $\lambda_m=0$
and a minimal value $0$ at $\lambda_m=1$,
while monotonically decreasing as $\lambda_m$ increases.
In fact,
measurement (\ref{eq:operator}) is
a projective measurement when $\lambda_m=0$
and is the identity operation when $\lambda_m=1$,
except for the unitary operation $\hat{U}_m$.


Unfortunately,
the measurement  changes the state of the qubit.
When the pre-measurement state is $\ket{\psi(a)}$
and the measurement outcome is $m$,
the post-measurement state is given by
\begin{equation}
   \ket{\psi(m,a)} = \frac{1}{\sqrt{p(m|a)}}\kappa_m\hat{U}_m\hat{D}_m
       \ket{\psi(a)}
\end{equation}
from Eqs.~(\ref{eq:reduction}) and (\ref{eq:operator}).
This state change can be quantified by the fidelity~\cite{Uhlman76,NieChu00}
between the pre-measurement and post-measurement states as
\begin{equation}
   F(m,a) = \bigl|\expv{\psi(a)|\psi(m,a)}\bigr|.
\end{equation}
As the process of measurement changes the state of qubit
to a greater extent, the fidelity becomes smaller.
Averaged over $a$ with the probability (\ref{eq:conditional}),
the fidelity after the measurement with the single outcome $m$
is evaluated as
\begin{equation}
   F(m) =\sum_a p(a|m)\bigl[F(m,a)\bigr]^2=\frac{1}{\,\overline{q_m}\,}
     \overline{\left|\bra{\psi}\hat{U}_m\hat{D}_m\ket{\psi}\right|^2}.
\label{eq:fide}
\end{equation}
Here, we have averaged the squared fidelity
rather than the fidelity for simplicity;
this choice does not qualitatively affect our results.
If the fidelity $F(m)$ is averaged over all the outcomes,
it reduces to the mean operation fidelity~\cite{Banasz01},
\begin{equation}
   F\equiv\sum_m p(m)\,F(m)=\sum_m
     \overline{\left|\bra{\psi}\hat{M}_m\ket{\psi}\right|^2}.
\label{eq:av_fide}
\end{equation}

To explicitly calculate the fidelity (\ref{eq:fide}),
we must specify the unitary operator $\hat{U}_m$,
in sharp contrast with the case of the information gain (\ref{eq:info}).
We parameterize it in the matrix representation as
\begin{equation}
 \hat{U}_m=e^{i\alpha_m}\left(\begin{array}{cc}
    e^{i\beta_m}\cos\gamma_m
    & -e^{i\delta_m}\sin\gamma_m \\
    e^{-i\delta_m}\sin\gamma_m
    & e^{-i\beta_m}\cos\gamma_m
    \end{array}\right),
\end{equation}
where $\alpha_m$, $\beta_m$, $\gamma_m$, and $\delta_m$ are real.
Therefore, the fidelity is calculated to be
\begin{equation}
   F(m)=\frac{1}{3}+\frac{1}{3}\left[1+\frac{2\lambda_m}{1+\lambda_m^2}
      \cos 2\beta_m \right]\cos^2\gamma_m.
\end{equation}
For a given $\lambda_m$, the lower and upper bounds
on the fidelity are given by
\begin{equation}
  \frac{1}{3} \le F(m) \le
   \frac{2}{3}\left[1+\frac{\lambda_m}{1+\lambda_m^2}\right].
\end{equation}
The lower bound does not depend on $\lambda_m$
and is achieved, e.g.,
if $\hat{U}_m=\ket{0}\bra{1}+\ket{1}\bra{0}$,
whereas the upper bound depends on $\lambda_m$
and is achieved, e.g., if $\hat{U}_m=\hat{I}$.
Because the unitary operator $\hat{U}_m$ causes the state change
irrelevant to the information gain $I(m)$,
the upper bound
\begin{equation}
  F_{\text{opt}}(m) \equiv
  \frac{2}{3}\left[1+\frac{\lambda_m}{1+\lambda_m^2}\right],
\label{eq:fidelity}
\end{equation}
which we refer to as optimal fidelity,
can be regarded as a measure of the inevitable state change
by the extraction of information through
the measurement operator $\hat{M}_m$.
The fidelity $F(m)$ is also shown in Fig.~\ref{fig1}
as a function of $\lambda_m$.
In particular,
the optimal fidelity $F_{\text{opt}}(m)$ has a minimal value
$2/3$ at $\lambda_m=0$
and a maximal value $1$ at $\lambda_m=1$,
while monotonically increasing as $\lambda_m$ increases.


Although the measurement changes the state of the qubit as mentioned above,
if the measurement is physically reversible~\cite{UeImNa96,Ueda97},
we can reverse this state change by a reversing measurement.
The reversing measurement is constructed so that
when it yields a preferred outcome (e.g., $0$,),
it applies a measurement operator
\begin{equation}
 \hat{R}_0^{(m)}= \eta_m \hat{M}_m^{-1}
   =\frac{\eta_m}{\kappa_m} \hat{D}_m^{-1}\hat{U}_m^\dagger
\end{equation}
with a complex number $\eta_m$
to the post-measurement state $\ket{\psi(m,a)}$ of the qubit,
thereby canceling the effect of $\hat{M}_m$ owing to
\begin{equation}
   \hat{R}_0^{(m)}\hat{M}_m=\eta_m \hat{I}.
\end{equation}
That is, when the reversing measurement on $\ket{\psi(m,a)}$
yields the preferred outcome $0$,
the state of the qubit reverts to the pre-measurement state $\ket{\psi(a)}$
except for an overall phase factor
via the state reduction (\ref{eq:reduction}),
\begin{equation}
  \ket{\psi_\text{rev}(m,a)}=
  \frac{1}{\sqrt{p_\text{rev}(m,a)}}\hat{R}_0^{(m)}\ket{\psi(m,a)}
  \propto \ket{\psi(a)},
\end{equation}
where $p_\text{rev}(m,a)$ is the successful probability
of the reversing measurement defined by
\begin{equation}
  p_\text{rev}(m,a) = \bra{\psi(m,a)}\hat{R}_0^{(m)^\dagger}\hat{R}_0^{(m)}
            \ket{\psi(m,a)}=\frac{|\eta_m|^2}{p(m|a)}
\end{equation}
as given in Eq.~(\ref{eq:probability}).
Here, we define the physical reversibility
by the maximal successful probability
of the reversing measurement~\cite{KoaUed99,Ban01,KorJor06}.
Since the upper bound on $|\eta_m|^2$ is given by~\cite{KoaUed99}
\begin{equation}
  |\eta_m|^2\le \inf_{\ket{\psi}}\, \bra{\psi}\hcs{\hat{M}_m}\ket{\psi}
   =\kappa_m^2\lambda_m^2
\end{equation}
to satisfy $\bra{\psi}\hat{R}_0^{(m)^\dagger}\hat{R}_0^{(m)}\ket{\psi}\le 1$
for any $\ket{\psi}$, the physical reversibility becomes
\begin{equation}
  R(m,a) \equiv \max_{\eta_m}\,p_\text{rev}(m,a)
   =\frac{\kappa_m^2\lambda_m^2}{p(m|a)}=\frac{\lambda_m^2}{q_m(a)}.
\end{equation}
Averaged over $a$ with the probability (\ref{eq:conditional}),
the reversibility of the measurement with the single outcome $m$
is evaluated as
\begin{equation}
  R(m) = \sum_a p(a|m)\,R(m,a)=\frac{\lambda_m^2}{\,\overline{q_m}\,}
   =\frac{2\lambda_m^2}{1+\lambda_m^2},
\label{eq:reversibility}
\end{equation}
which depends only on $\lambda_m$.
The reversibility $R(m)$ is also shown in Fig.~\ref{fig1}
as a function of $\lambda_m$.
It has a minimal value $0$ at $\lambda_m=0$
and a maximal value $1$ at $\lambda_m=1$,
while monotonically increasing as $\lambda_m$ increases.
Clearly, measurement (\ref{eq:operator}) is physically reversible
unless $\lambda_m= 0$.
If the reversibility $R(m)$ is averaged over all the outcomes,
it reduces to the degree of physical reversibility of measurement
discussed by Koashi and Ueda~\cite{KoaUed99},
\begin{equation}
   R\equiv\sum_m p(m)\,R(m)
    =\sum_m \inf_{\ket{\psi}}\, \bra{\psi}\hcs{\hat{M}_m}\ket{\psi}.
\label{eq:av_reve}
\end{equation}

\section{\label{sec:tradeoff}Tradeoff Relations}
\begin{figure}
\begin{center}
\includegraphics[scale=0.6]{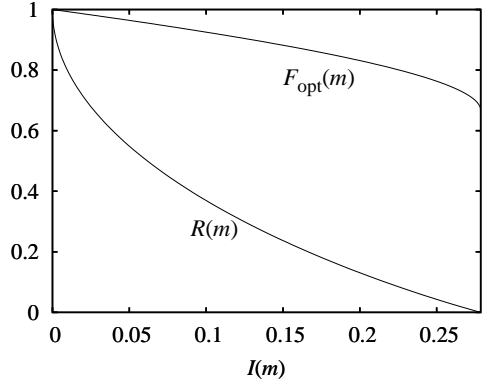}
\end{center}
\caption{\label{fig2}
Optimal fidelity $F_{\text{opt}}(m)$ and reversibility $R(m)$
as functions of the information gain $I(m)$.}
\end{figure}

Since we have written
the information gain $I(m)$, the optimal fidelity $F_{\text{opt}}(m)$,
and the reversibility $R(m)$ as functions of
the same single parameter $\lambda_m$,
as given in Eqs.~(\ref{eq:information}),
(\ref{eq:fidelity}), and (\ref{eq:reversibility}), respectively,
it is easy to find relations among them.
In fact, we can plot $F_{\text{opt}}(m)$ and $R(m)$
as functions of $I(m)$, as in Fig.~\ref{fig2},
to show trade-off relations
\emph{at a single outcome level}.
That is, as the measurement provides more information about
the state of the qubit,
the process of measurement changes the state to a greater extent
and makes it even less reversible.
\begin{figure}
\begin{center}
\includegraphics[scale=0.6]{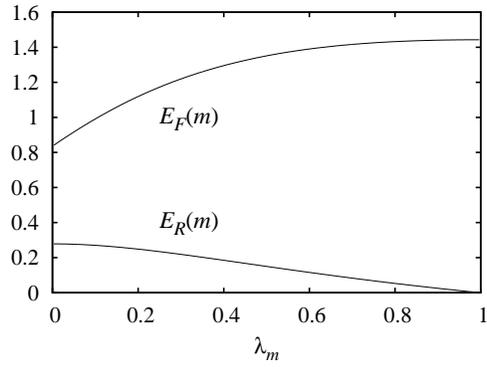}
\end{center}
\caption{\label{fig3}
Efficiencies $E_F(m)$ and $E_R(m)$ of measurement
as functions of $\lambda_m$.}
\end{figure}
These tradeoff relations derive
two types of measurement efficiencies:
the ratio of the information gain to the optimal fidelity loss
\begin{equation}
    E_F(m)\equiv \frac{I(m)}{1-F_{\text{opt}}(m)}
\end{equation}
and the ratio of the information gain to the reversibility loss
\begin{equation}
    E_R(m)\equiv \frac{I(m)}{1-R(m)}.
\end{equation}
Figure~\ref{fig3} shows the efficiencies $E_F(m)$ and $E_R(m)$
as functions of $\lambda_m$.
Note that
$E_F(m)$ is a monotonically increasing function,
whereas $E_R(m)$ is a monotonically decreasing function.
Therefore,
at $\lambda_m=0$,
$E_F(m)$ has a minimal value $3[1-1/(2\ln2)]$
and $E_R(m)$ has a maximal value $1-1/(2\ln2)$.
This means that
the projective measurement, which provides the most information
and causes the largest state change with no reversibility,
is the most efficient with respect to the reversibility
but is the least efficient with respect to the fidelity.
In the limit of $\lambda_m\to 1$,
we obtain $E_F(m)\to 1/\ln 2$ and $E_R(m)\to 0$.

\section{\label{sec:conclude}Conclusion}
In conclusion,
we calculated the information gain, fidelity,
and physical reversibility of an arbitrary single-qubit measurement,
assuming that the qubit to be measured was in a completely unknown state.
These quantities are expressed as functions of
the same single parameter $\lambda_m$,
which is the ratio of the two singular values of the measurement operator
corresponding to the outcome,
as shown in Eqs.~(\ref{eq:information}),
(\ref{eq:fidelity}), and (\ref{eq:reversibility}).
Our results gave information tradeoff relations
to the fidelity and reversibility
at the level of a single outcome
without averaging all outcomes.
Moreover, two efficiencies of the measurement were discussed
to show their different behaviors:
the ratio of the information gain to the optimal fidelity loss
and the ratio of the information gain
to the reversibility loss.
As the information gain decreases by increasing
the parameter $\lambda_m$,
the former ratio increases
whereas the latter decreases.

Our tradeoff relations are
applicable to any efficient measurement
on a qubit or two-level system
with postselection.
A characteristic feature of our tradeoff relations is
that the information gain is directly
related to the fidelity and reversibility
for a given measurement $\hat{M}_m$,
because all the quantities are functions of
the single parameter $\lambda_m$.
By only eliminating the parameter $\lambda_m$,
we can obtain the tradeoff curves, as shown in Fig.~\ref{fig2},
without optimization problems~\cite{Banasz01,BanDev01,Sacchi06}.
Unfortunately, this does not apply to
more general situations.
For example, in measurements with an overall outcome average,
the information gain (\ref{eq:av_info}),
fidelity (\ref{eq:av_fide}), and
reversibility (\ref{eq:av_reve})
are functions of all $\{\lambda_m\}$ and
$\{\kappa_m\}$ corresponding to possible outcomes
because $m$ is summed over by using the value of
the total probability (\ref{eq:totalprob}),
\begin{equation}
  p(m)=\frac{1}{2}\kappa_m^2\left(1+\lambda_m^2\right).
\end{equation}
In measurements on $d$-level systems
such as qudit or multiple qubits,
all quantities are functions of $d-1$ parameters
$\{\lambda_m^{(1)},\ldots,\lambda_m^{(d-1)}\}$
because the measurement operator is represented by
a $d\times d$ matrix in an orthonormal basis.
Moreover, in measurements with classical noise,
they are functions of multiple parameters
because a single measurement process
is described by a set of measurement operators.
To find tradeoff curves in such situations,
we must optimize measurements by maximizing
the fidelity or the reversibility
with a fixed value of the information gain by
using numerical calculations.
Our simple and direct tradeoff relations are
free from such optimization problems;
therefore, they can be
regarded as highly fundamental in quantum measurement.

\section*{Acknowledgments}
This research was supported by a Grant-in-Aid
for Scientific Research (Grant No.~20740230) from
the Ministry of Education, Culture, Sports,
Science and Technology of Japan.


\end{document}